\documentclass[useAMS,usenatbib]{mn2e}
\usepackage{mathptmx}
\usepackage{graphicx}
\usepackage{lscape}
\usepackage{listings}
\title[LOFAR brown dwarfs]{A LOFAR mini-survey for low-frequency radio emission from the nearest brown dwarfs}
\author[Burningham et al]{Ben Burningham$^{1,2}$\thanks{E-mail:
    B.Burningham@herts.ac.uk}, M. Hardcastle$^{2}$, J. D. Nichols$^{3}$, S.L. Casewell$^{3}$, S.P. Littlefair$^{4}$, 
\newauthor
C. Stark$^{5}$, M. R. Burleigh$^{3}$,S. Metchev$^{6,7}$, M. E. Tannock$^{6}$, R. J. van Weeren$^{8}$, 
\newauthor
W. L. Williams$^{2}$, G. A. Wynn$^{3}$
\\   
$^{1}$ NASA Ames Research Center, Mail Stop 245-3, Moffett Field, CA 94035, USA \\
$^{2}$ Centre for Astrophysics Research, School of Physics, Astronomy and Mathematics, University of Hertfordshire, Hatfield AL10 9AB \\
$^{3}$ Department of Physics and Astronomy, University of Leicester, Leicester LE1 7RH, UK\\
$^{4}$ Department of Physics and Astronomy, University of Sheffield, Sheffield S3 7RH, UK\\
$^{5}$ Division of Computing and Mathematics, Kydd Building, Abertay University, Dundee DD1 1HG, UK\\
$^{6}$ Centre for Planetary Science and Exploration, Department of Physics \& Astronomy, The University of Western Ontario, London, ON N6A 3K7, Canada\\
$^{7}$ Department of Physics \& Astronomy, Stony Brook University, Stony Brook, NY 11794-3800, USA \\
$^{8}$ Harvard-Smithsonian Center for Astrophysics, 60 Garden Street,
  Cambridge, MA 02138, USA
}

\begin{document}
%
%
%
%


\def\aj{\rm{AJ}}                   
\def\araa{\rm{ARA\&A}}             
\def\apj{\rm{ApJ}}                 
\def\apjl{\rm{ApJ}}                
\def\apjs{\rm{ApJS}}               
\def\ao{\rm{Appl.~Opt.}}           
\def\apss{\rm{Ap\&SS}}             
\def\aap{\rm{A\&A}}                
\def\aapr{\rm{A\&A~Rev.}}          
\def\aaps{\rm{A\&AS}}              
\def\azh{\rm{AZh}}                 
\def\baas{\rm{BAAS}}               
\def\jrasc{\rm{JRASC}}             
\def\memras{\rm{MmRAS}}            
\def\mnras{\rm{MNRAS}}             
\def\pra{\rm{Phys.~Rev.~A}}        
\def\prb{\rm{Phys.~Rev.~B}}        
\def\prc{\rm{Phys.~Rev.~C}}        
\def\prd{\rm{Phys.~Rev.~D}}        
\def\pre{\rm{Phys.~Rev.~E}}        
\def\prl{\rm{Phys.~Rev.~Lett.}}    
\def\pasp{\rm{PASP}}               
\def\pasj{\rm{PASJ}}               
\def\qjras{\rm{QJRAS}}             
\def\skytel{\rm{S\&T}}             
\def\solphys{\rm{Sol.~Phys.}}      
\def\sovast{\rm{Soviet~Ast.}}      
\def\ssr{\rm{Space~Sci.~Rev.}}     
\def\zap{\rm{ZAp}}                 
\def\nat{\rm{Nature}}              
\def\iaucirc{\rm{IAU~Circ.}}       
\def\aplett{\rm{Astrophys.~Lett.}} 
\def\apspr{\rm{Astrophys.~Space~Phys.~Res.}}
\def\bain{\rm{Bull.~Astron.~Inst.~Netherlands}} 
\def\fcp{\rm{Fund.~Cosmic~Phys.}}  
\def\gca{\rm{Geochim.~Cosmochim.~Acta}}   
\def\grl{\rm{Geophys.~Res.~Lett.}} 
\def\jcp{\rm{J.~Chem.~Phys.}}      
\def\jgr{\rm{J.~Geophys.~Res.}}    
\def\jqsrt{\rm{J.~Quant.~Spec.~Radiat.~Transf.}}
\def\memsai{\rm{Mem.~Soc.~Astron.~Italiana}}
\def\nphysa{\rm{Nucl.~Phys.~A}}   
\def\physrep{\rm{Phys.~Rep.}}   
\def\physscr{\rm{Phys.~Scr}}   
\def\planss{\rm{Planet.~Space~Sci.}}   
\def\procspie{\rm{Proc.~SPIE}}   

\let\astap=\aap
\let\apjlett=\apjl
\let\apjsupp=\apjs
\let\applopt=\ao

\maketitle
%
%

\begin{abstract}
We have conducted a mini-survey for low-frequency radio emission from some of the closest brown dwarfs to the Sun with rapid rotation rates: SIMP~J013656.5 +093347, WISEPC~J150649.97+702736.0, and WISEPA~J174124.26+255319.5. 
We have placed robust 3$\sigma$ upper limits on the flux density in the 111 -- 169 MHz frequency
range for these targets: WISE 1506: $< 0.72$~mJy; WISE 1741: $< 0.87$~mJy; SIMP 0136: $< 0.66$~mJy.  At 8 hours of integration per target to achieve these limits, we find that systematic and detailed study of this class of object at LOFAR frequencies will require a substantial dedication of resources. 

\end{abstract}


\section{Introduction}
\label{sec:intro}

With effective temperatures ranging from 2400~K to below 400~K, the
atmospheres of ultracool dwarfs (UCDs) link the stellar and planetary
regimes. The warmest UCDs extend the cool stellar sequence from the
spectral type M7 to the L spectral class ($2400~K > T_{\rm eff} > 
1400~K$), incorporating objects of both stellar and substellar
masses, and displaying optical and infrared spectral morphologies
shaped by the development of thick silicate condensate clouds \citep[e.g. ][]{kirkpatrick1999}. 
At cooler temperatures ($1400~K > T_{\rm eff} > 300~K$), T~and Y~dwarfs are
exclusively substellar,  with the spectral sequence defined by the development of broad absorption bands due to water, methane and ammonia \citep[e.g. ][]{burgasser2006, cushing2011,canty2015,line2015}.
The defining feature of brown dwarfs is their failure to reach the
main sequence, meaning that they cool and fade as they age. As a
result, brown dwarfs of planetary mass (a few $M_{Jupiter}$) have been found across the L, T and Y
dwarf spectral classes at a variety of ages and temperatures \citep[e.g. ][]{burgasser2010,ben2011b,cushing2011, faherty2013,liu2013,delorme2013,naud2014}. 
Spanning this transitionary parameter space, substellar UCDs represent outstanding opportunities for understanding the physics,
chemistry and dynamics that differentiate the stellar and planetary domains.
This is particularly true for understanding magnetic fields and their interaction with atmospheres and space-environments.

Fully convective early type M dwarfs display well documented evidence
of stellar magnetic 
activity in the form of flares, and correlated X-ray, H$\alpha$ and
radio emission.  However, the fraction of total luminosity emitted via optical and X-ray tracers of activity drops off
rapidly in the UCD regime as atmospheric ion fractions drop with decreasing
temperature, and as the increasingly neutral atmospheres decouple from
magnetic fields \citep{gizis2000,mohanty2002}.
Surprisingly,  radio luminosity does not drop off, thereby violating the correlation between radio and X-ray flux typically seen in stellar sources \citep{gudel1993,benz1994}. 
This fact, and the observed brightness temperatures of the radio emission, suggests that despite the cool temperatures, a significant source of plasma exists in the envelope surrounding the UCD radio sources. 
Radio emission has been detected in $\approx 5 - 10 \%$ of UCDs across spectral types ranging from M9 to T6, spanning a $T_{\rm eff}$ range of some 1500~K, incorporating largely neutral photospheres \citep[e.g. ][]{berger2001, berger2006,mclean2012, route2012, williams2013,kao2016,lynch2016}, although the majority of these have earlier spectral types. Indeed, until the survey of \citet{kao2016},  over 60 objects with spectral types later than L6 had been surveyed, but only one has been detected to be radio emitting \citep{antonova2013, route2013}. Characterising this new population of radio emitters is now the target of significant enquiry \citep[e.g. ][]{mclean2012,williams2013, kao2016}.

Pulsed periodic and quiescent emission has been detected from the currently identified radio `active' UCDs via a number of studies in the 8.5 GHz and 4.9 GHz radio bands \citep[e.g][]{hallinan2006,hallinan2007, hallinan2008, berger2009, williams2014}. The periodic pulses are coherent, apparently rotationally modulated and typically display 100\% polarisation. They have been attributed to the electron cyclotron maser instability \cite[ECMI; e.g. ][]{treumann2006, hallinan2008}. Quiescent emission has been detected in all cases where pulsed emission is present,  and is essentially unpolarised. It has been explained as either a component of the ECMI emission that has been depolarised during transmission through the UCD's magnetosphere \citep{hallinan2008} or as gyrosynchrotron emission \citep{berger2002}.
Gyrosynchrotron emission has been suggested as the cause of non-flaring emission detected in an ultracool dwarf (TVLM513-46546) at 95 GHz 
with the Atacama Large Millimeter Array \citep{williams2015}. This object also emits via the ECMI mechanism (which couldn't produce 95 GHz emission for plausible field strengths), emitting periodic pulses on its rotation period.  \citet{ravi2011} also suggest gyrosynchrotron emission is responsible for non-flaring $4.5 - 24$~GHz radio emission detected from DENIS-P~J104814.9--395604. TVLM513-46546 was also observed at 325~Mhz without detection at 0.8 mJy level the  by \citet{jaeger2011}.

The ECMI mechanism is responsible for radio emission across a variety of astronomical contexts, from Solar system system planets to emission from some types of solar flare through to massive star radio emission \citep{treumann2006}. ECMI emission is inherently variable, and produces directional (beamed) circularly- or elliptically-polarised emission, consistent with the rotationally modulated pulses seen in UCDs. 
Recently, \citet{nichols2012} have modelled the ECMI emission from UCDs as a originating from the upward magnetic field-aligned component of a large-scale magnetosphere-ionosphere coupling current system flowing as a result of a meridional angular velocity gradient in ionospheric plasma. The \citet{nichols2012} model broadly matches the properties of the emission seen in the UCDs to which it has been applied.  Furthermore, the wavelength-dependence of the optical variability of radio-emitting UCDs has been interpreted as arising as a result of non-thermal ionisation caused when the electron currents impact the photosphere \citep{hallinan2015}. Thus, the optical and radio emission can be seen as evidence for powerful auroral currents in UCDs, similar to e.g.\ Jupiter's main auroral oval \citep{grodent03} and significant components of its decametric, hectometric and kilometric radio bursts \citep{zarka98}.

At Jupiter, the flow shear is a result of centrifugally-driven outward diffusion of plasma generated at the volcanic moon Io, which orbits deep within the rapidly rotating jovian magnetosphere \citep{hill79,cowley01,nichols03,nichols04,nichols05}.  Conservation of angular momentum results in a gradient in the angular velocity with radial distance, which, when mapped along magnetic field lines to the planet, results in a meridional current flowing in the Pedersen conducting layer of the atmosphere (i.e.\ where the collisional frequency is comparable to  the gyrofrequency, allowing maximum ion mobility in the direction of the electric field).  From current continuity, the divergence of the Pedersen current yields the field-aligned current, the upward component of which (corresponding to downward-precipitating electrons) gives rise to the auroral and radio emissions.  \cite{nichols2012} hypothesised that either such centrifugally-driven outflow or the interaction of the rotating magnetosphere with an external flowing medium \citep{isbell84} could produce an angular velocity gradient at brown dwarfs.  Regardless of the ultimate cause of the flow shear, however, the expected radio spectrum is essentially flat, with a high-frequency cut-off set by the cyclotron frequency $\nu = eB_i / 2\pi m_e$ (where $B_i$ is the polar magnetospheric magnetic field strength) at the top of the ionosphere, and a low frequency cut-off given by the cyclotron frequency at the altitude of the field-aligned voltage that drives the upward field-aligned current.  The radio luminosity is principally dependent on the dwarf's angular velocity, magnetic field strength, ionospheric Pedersen conductance, and the properties of the high latitude plasma, so its measurement provides the opportunity to study the plasma environment of brown dwarfs and identify departures and similarities to jovian case, such as (for example) the presence or absence of an Io-analogue companion. 

Despite the progress made in the field, a number of  problems remain for understanding the radio properties of the wider population.  Most pressing is identifying which properties distinguish the radio `active' objects from the `inactive' ones. 
Overall, about 5\% of UCDs have been detected at GHz bands, and the hit rate is biased towards types earlier than L3.5, for which the detection rate is $\sim 10\%$ compared with $\sim 2\%$ for types later than L3.5 \citep{antonova2013}.

Several explanations could explain this low rate of detection. Firstly, the electrodynamic engine responsible for the radio emission may be absent or too weak to power detectable emission in the majority of cases. 
If the electrodynamic engine is co-rotation breakdown as in the case of the Jovian main auroral oval and the \citet{nichols2012} model, then this may indicate e.g. a lack of plasma sources, small ionospheric Pedersen conductance or insufficient velocity shear. Similarly, if the engine is the interaction between the UCD magnetic field and orbiting planets then the low-detection may simply indicate the rarity of such systems.

Another possibility is that geometric effects might hide radio emitters, since the beaming of the ECMI generated emission can be expected to restrict the visibility to only favourable geometries. However, the low detection rates would require a much tighter typical beam than is seen for Jupiter's auroral radio emission ($\approx 1.6$~sr), which would give rise to an expected detection rate of about 46\% in the absence of other factors.
 The pulse duty cycles, which range from 0.05 to 0.3, also suggest that beaming is unlikely to account for the low detection rate. This is further supported by the results of \citet{pineda2016}, who find that the H$\alpha$ detection rate in this regime is similar to the radio detection rate. The dichotomy in detection rates  between objects earlier than L3.5 and later types further suggests that geometry is not the dominant factor. In addition, the quiescent emission is often only a factor of a few fainter than the pulses \citep[e.g.][]{hallinan2008}, and it should be unaffected by viewing angle. 

The remaining likely cause for the low-detection rate is the fact that all UCD radio surveys to-date require local magnetic field strengths above 1.6~kG, due to the high-frequency cut-off characteristic of ECMI generated emission. The capability of LOFAR to access the MHz frequency domain allows us to target objects with magnetic field strengths of tens of Gauss. 
To establish the feasibility of using LOFAR to study the auroral environment of cool brown dwarfs with sub-kG fields, we have conducted a mini-survey of nearby brown dwarfs.

\section{Target selection}
\label{sec:sel}

We selected three targets for deep LOFAR imaging based on proximity to the Solar system, and (where available) short rotation period.  The aim was not to create an unbiased survey, but to select targets which maximised the chances of detecting low-frequency emission in the \citet{nichols2012} model. We targeted three nearby brown dwarfs comprising one early-T~dwarf: SIMP J013656.5 +093347 \citep[hereafter SIMP0136; ][]{artigau2006}; and two late-T~dwarfs: WISEPC J150649.97+702736.0, \citep[hereafter WISE1506; ][]{kirkpatrick2011}, and WISEPA J174124.26+255319.5 \citep[hereafter WISE1741; ][]{scholz2011,gelino2011,kirkpatrick2011}. Our target properties are summarised in Table~\ref{tab:targets}.

\begin{table*}
\begin{tabular}{c c c c c c c}
\hline
Target & Spectral Type & D (pc) & $P_{rot}$ (hours) & Mass ($M_{Jupiter}$) & $T_{\rm eff}$ (K) & refs \\
\hline
SIMP~0136 & T2.5 & $6.4 \pm 0.3$ & 2.5 & 40 -- 70 & $1300 \pm 200$ & 1,1,1,e,e \\
WISE~1506 & T6 & $3.4 ^{+0.7}_{-0.4}$ & 1.74 & 30 -- 60 & $1000 \pm 200$ & 2,3,4,e,e\\
WISE~1741 & T9 & $5.6^{+0.5}_{-0.4}$ & - & 10 -- 35 & $620 \pm 60$ & 2,5,-,5,5\\
\hline
\end{tabular}
\caption{\label{tab:targets}Summary of properties for our targets. Refs: `e' indicates a rough estimate based on spectral type for $T_{\rm eff}$ and assuming and age range of 1 -- 10~Gyr for the adopted mass. Otherwise, number indicates literature values taken from: 1) \citet{artigau2006};2) \citet{kirkpatrick2011}; 3) \citet{marsh2013}; 4) Tannock et al (in prep);  5) \citet{dupuy2013}.}
\end{table*}

It should be noted that SIMP0136 was recently studied in detail by \citet{apai2013} who used time-resolved spectroscopy to map the surface of the brown dwarf. SIMP0136 appears to be variable on its rotational timescale which has been attributed to a patchy atmosphere, with the surface covered by two distinct regions consisting of low-temperature thick clouds, and higher temperature, thin, bright clouds.  Following the results of \citet{hallinan2015} linking photometric variability and radio emission to aurorae,  \citet{kao2016} performed a survey targeting ultracool dwarfs with known optical/near-IR variability. They detected at least one circularly polarised radio pulse from SIMP0136 in the 4-8 GHz regime, as well as quiescent emission, and determined the surface $B$ field strength to be a minimum of 2.5 kG. Neither of the other two targets have been detected in the radio. 

We also note that since our targets were observed, SIMP~0136 and WISE~1506 were observed as part of the \citet{pineda2016} brown dwarf H$\alpha$ survey, and neither showed H$\alpha$ emission, with limits of $f_{\alpha} < 4.9 \times 10^{-18}$ ergs~s$^{-1}$~cm$^{-2}$ and $f_{\alpha} < 5.8 \times 10^{-18}$ ergs~s$^{-1}$~cm$^{-2}$ reported respectively. 

In addition, WISE 1506 was monitored photometrically with Spitzer as part of GO program 11174 (PI: S. Metchev) to seek cloud-induced variability in a second phase of the Weather on Other Worlds survey (Metchev et al. 2015, ApJ, 799, 154).  WISE 1506 shows periodic variability in these observations with a period of 1.74 hr (Tannock et al., in prep.).

\section{Observations and image construction}
\label{sec:obs}
We observed the targets in the frequency range between 111 and 169 MHz
at 64 channels per sub-band in 8-bit mode using the HBA\_DUAL\_INNER
configuration, with a standard integration time of 1 s. The
integration time and frequency binning were designed to facilitate RFI
removal. Each target was observed for a total of approximately 8
hours, but because of the low declination of SIMP0136 the observing
time was broken up into two observations on two consecutive days (see
Table \ref{tab:obs} for details). In addition to providing the obvious
benefit of a long integration, the observation duration was chosen to
ensure that at least one rotation period was observed for all targets.
In each case, the observations of the target field were preceded and
followed by short, 10-minute observations of flux calibrator sources.

 \begin{table*}
  \caption{Observational details for the three targets}
  \label{tab:obs}
  \begin{tabular}{lrrrrr}
    \hline
    Target&Start date&Start time&End date&End time&Time on source (h)\\
    \hline
    WISE1506 & 2014-07-24 & 16:16:00 & 2014-07-25 & 00:16:00 & 8.00\\
    SIMP0136 (1)&2014-08-17 & 01:16:00 & 2014-08-17 & 05:01:09 & 3.75\\
    SIMP0136 (2)&2014-08-18 & 01:16:00&2014-08-18&05:01:09&3.75\\
    WISE1741 & 2014-10-01 & 13:16:00 & 2014-10-01 & 21:16:00 & 8.00\\
    \hline
  \end{tabular}
  \end{table*}

After observation, the data were averaged by the observatory to 4
channels per sub-band (an HBA sub-band has a bandwidth of 200 kHz) and
a 5-second integration time. No `demixing' of bright off-axis sources
was carried out by the observatory -- this was deemed unnecessary
given the sky positions of bright objects like Cygnus A and Cas A --
and all further processing was carried out by us using the University
of Hertfordshire high-performance computing facility.

The data were processed using the `facet calibration' techniques which
are described in detail by \citet[][; hereafter vW16]{vanWeeren+16}
and \citet{williams2016}, with the preliminary processing implemented
by us on the Hertfordshire system as described by \citet[][ H16]{hardcastle2016}. 
Here we provide only a very brief overview of the
data processing, and the reader is referred to vW16 and H16 for more
details. After flagging for RFI and correcting for the effects of
remote station clock offsets, the data were combined into `bands' of
ten sub-bands each (with a total bandwidth of just under 2 MHz) and
averaged again for computational speed to a 10-s integration time and
20 channels per band. As noted by H16, these parameters result in
  bandwidth and time-averaging smearing that affect peak flux
  densities at the $\sim 10$ per cent level in full-resolution imaging
  at 2-3 degrees from the pointing centre, but our aim in imaging the
  full LOFAR field of view for this project is simply to obtain the
  best possible sensitivity at the pointing centre; the averaging
  applied has no effect in the vicinity of our brown dwarf targets.
The data were then phase self-calibrated using a model derived from
combining the NRAO VLA Sky Survey (NVSS) with the LOFAR 150-MHz image
to give initial phase solutions, and this provided phase solutions
good enough to allow imaging of each band and subtraction of all the
detected sources, giving the blank (residual) images and sky models
needed for the facet calibration process.

Our facet calibration process differs from that described by vW16 in
that we did not use the whole LOFAR bandwidth for the fitting, but
instead divided the source into five spectral windows each consisting
of 6 2-MHz bands. As discussed by H16, this process has both
advantages and disadvantages when compared to the vW16 approach. The
advantages are (i) that each spectral window can be calibrated in
parallel, speeding up the process considerably; (ii) that flux scale
corrections can be derived as a function of frequency rather than
across the whole LOFAR band, as described below; and (iii) that we can
measure flux density across the band, potentially allowing the
measurement of in-band spectral index. The disadvantages are that
there is less signal to noise for the self-calibration required in the
facet calibration process, potentially making the results less good,
that phase offsets may be introduced between the different datasets,
and that since the resulting images must be combined in the image
plane, requiring us to convolve all images to the same resolution
  before summing them, the image resolution is limited to the
resolution of the lowest-frequency LOFAR band. In the present case, we
required good flux calibration and wished to be able to constrain the
source spectrum in the event of a BD detection, so we considered the
advantages to outweigh the disadvantages. Accordingly, we used the
five spectral windows listed in Table~\ref{tab:spw}, giving a total of
about 60 MHz of bandwidth. The upper end of the HBA bandwidth is badly
affected by RFI and was not used.
 
 \begin{table}
   \caption{Spectral windows, frequencies and LOFAR band/sub-band
    numbers used in facet calibration}
   \label{tab:spw}
\begin{tabular}{rrrr}
\hline
    Spectral&Frequency range&Band&Sub-band\\
    window&(MHz)&numbers&numbers\\
    \hline
    1&111-122&0-5&0-59\\
    2&122-134&6-11&60-119\\
    3&134-146&12-17&120-179\\
    4&146-157&18-23&180-239\\
    5&158-169&24-29&240-299\\
    \hline
  \end{tabular}
\end{table}

The facet calibration process was run on these five spectral windows,
imaging the whole $\sim 30$-deg$^2$ LOFAR field of view. This gives us
for each target five large (approx $20,000 \times 20,000$ 1.5-arcsec
pixels) images, one for each spectral window (we combined the two
observations of SIMP 0136 in the image plane at this point). The
images at each frequency band were convolved to matching resolution
and a primary beam correction was applied in the image plane.

Finally, we had to correct the flux scale for these images to remove
the frequency-dependent effects of the fact that transfer of the gains
from the flux calibrator does not take full account of the effect on
the beam of the different elevations of calibrator and target source.
We did this (see H16 for more details) by extracting a multi-frequency
source catalogue for each image using {\sc pybdsm}, cross-matching
with the VLSSr and NVSS surveys to produce a small sample of $\sim 50$
bright sources detected in all bands, and determining the scaling
factors for each spectral window that produced the best power-law fits
to the spectra of all sources in the source catalogue after outlier
rejection. Total LOFAR flux densities are used in this process,
  so it is unaffected by bandwidth or time-averaging smearing at large
  distances from the pointing centre; only sources that have
  deconvolved angular sizes less than 20 arcsec in the LOFAR images,
  and are thus unresolved in NVSS and VLSS images, are used to avoid
  any effects due to the mismatch in LOFAR and NVSS/VLSS resolutions.
With these correction factors applied, the flux densities we measure
should be correct on the scale of \cite{Scaife+Heald12}, since both
VLSSr and NVSS are on this scale. A caveat is that there may well be
some spectral curvature between the LOFAR and NVSS frequencies, which
would tend to cause us to underestimate the magnitude of the flux
scale correction, particularly at the high-frequency end of the LOFAR
band. While this is probably a real effect, it can only be overcome if
a lower-frequency survey than NVSS is available, which is not the case
for all of our fields. We chose to use NVSS for all three fields for
consistency. The nominal statistical error on the correction
  factors is of order 1 per cent, but systematic effects, including
  that of spectral curvature, are probably larger; the limiting factor
  is probably the VLSSr absolute calibration, which we estimate to be
  good to $\sim 10$ per cent based on the discussion by
  \citet{Lane+14}. For the purposes of this paper, 10 per cent flux
  density uncertainties are acceptable.

With the correction factors applied, we combined the images for the
five spectral windows to produce a single image at a nominal frequency
of 140 MHz, with a total bandwidth of 58 MHz.
 
\section{Results}
\label{sec:res}

Images of the fields around the brown dwarfs are presented in Appendix
A.

The typical central rms noise in our 140-MHz images is a little
over 200 $\mu$Jy beam$^{-1}$. This is about a factor 2 worse than the
best fields of vW16 and H16. We consider this to be the result of a
combination of factors. Facet calibration worked well in the SIMP0136
field, but on the equator the sensitivity of LOFAR is necessarily
lower because of projection effects. The WISE1506 field contains two
very bright sources, 3C\,309.1 and 3C\,314.1, and it is challenging to
remove these completely even with facet calibration, leading to a
higher level of noise. And WISE1741 lies close to a bright ($\sim 0.8$
Jy), very extended (11 arcmin) extragalactic source with complex
structure which we were unable to deal with well in facet calibration,
increasing the noise in the facet containing the target.

No sources are visible at (or near) the locations of any of the brown
dwarf targets. We estimate the following 3$\sigma$ detection limits
for our targets integrated across the full 111 -- 169 MHz frequency
range: WISE 1506: $< 0.72$~mJy; WISE 1741: $< 0.87$~mJy; SIMP 0136: $<
0.66$~mJy.

\section{Discussion}
\label{sec:disc}

A key aim of these observations was to assess whether the low-detection rate of brown dwarfs at GHz frequencies was due to local magnetic field strengths being sufficiently low to restrict ECMI emission to the 100 MHz frequency regime. Our lack of detection for any of our targets does not provide a basis for a strong conclusion in this matter.

We note that SIMP~0136 has been detected by \citet{kao2016} with a pulse spectral flux density of 0.23~mJy at 4-8~GHz. Assuming a flat spectrum, this would place this target below our detection limits.
This implies that the radio luminosity of this object does not increase significantly with decreasing frequency. 
 The GHz observations of SIMP~0136 by \citet{kao2016} reached a depth $\sim$100 times fainter than our LOFAR observations, and it is reasonable to infer that an increase in sensitivity at this level will be required to robustly test for a higher detection fraction at low-frequency.  
Moreover, such sensitive observations of GHz detected objects could also set useful constraints on the location of the field-aligned voltage driving the currents. For example, if it is located at an altitude less than $\sim$2.5 stellar radii up the dipolar field lines, then no emission would be expected at LOFAR frequencies.

The GHz detection of SIMP~0136 tells us that the electrodynamic engine that drives the radio emission is present in this case, and we are thus able to derive useful information from our upper limits on the low-frequency flux density in light of this fact. However, in the other two cases any further interpretation of our results relies on the assumption that the presumed electrodynamic engine is present.  Since the nature of the electrodynamic engine for the auroral emissions in brown dwarfs is not determined, this is highly speculative and the following discussion in respect of these two targets must be considered accordingly.

By considering the theoretical framework outlined by \cite{nichols2012}, we now use the upper limits to the spectral flux densities derived here to estimate upper limits of the Pedersen conductance of these bodies, and again highlight the caveats discussed in the previous paragraph.  We employ the fiducial parameters considered by \cite{nichols2012}, i.e.\ angular velocities which transition from 25\% to full corotation over the space of $\sim$1$^\circ{}$ centred on 15$^\circ{}$ co-latitude, as shown in their Fig. 1a, and values for the high latitude electron population temperature and number density of 2.5~keV and 0.01~cm$^{-3}$, respectively, i.e.\ jovian values.  We recognise that these parameters are unknown at brown dwarfs, and thus may differ considerably to these figures but in the absence of data to the contrary, the jovian figures enable a reasonable initial impression as to what conductances might be expected (note the expected power variation in respect to these quantities is shown in  Fig. 2 of \cite{nichols2012}).  

In addition to assuming a jovian-like plasma environment, we adopt the lower limit estimated by \citet{kao2016} of 2.5~kG as our polar magnetospheric magnetic field strength, $B_i$, for SIMP0136. For the other two targets we estimate their polar magnetic field strengths following \citet{reiners10}. For WISE1741 we adopt a rotation period at the lower end of the so-far observed $P_{rot}$ distribution of brown dwarfs \citep[e.g.][]{metchev2015} since we are seeking a upper limit on the Pedersen conductance.

Employing these parameter values, along with those indicated in Table~\ref{tab:sig} and assuming that the radio is beamed into the canonical 1.6~sr yields upper limits for the Pedersen conductances of order a tenth of a mho in each case, i.e.\ consistent with the lower end of jovian estimates, which range from $\sim 0.1$~mho to $\sim 10$~mho \citep[e.g.][]{strobel83,bunce2001}.

\begin{table}
\centering
\caption{Apopted parameters for our UCDs and resulting Pedersen conductance estimates.  Please see caveats in Section~\ref{sec:disc}.}
\label{tab:sig}
\begin{tabular}{l|lll}\hline
Target        & $B_i$ / kG & $P_{rot}$ / hours & $\Sigma_P$ / mho \\ \hline
SIMP 0136 & 2.5   & 2.5       & $<0.3$        \\
WISE 1506 & 0.7   & 1.74      & $<0.1$        \\
WISE 1741 & 0.3   & 2         & $<0.5$       \\
\hline
\end{tabular}
\end{table}

\section{Conclusions}
\label{sec:conc}
We have performed a limited search for low-frequency radio emission from some of the nearest rapidly rotating brown dwarfs to the Sun. We did not detect any of them, and estimate the following 3$\sigma$ detection limits
for our targets integrated across the full 111 -- 169 MHz frequency
range: WISE 1506: $< 0.72$~mJy; WISE 1741: $< 0.87$~mJy; SIMP 0136: $<
0.66$~mJy.  These limits are consistent with these objects displaying ionospheric Pedersen conductances of similar magnitude to those found in the jovian environment or lower. WISE 1506, in particular, is at the lower extreme of the jovian scenario. 

The significant observational time required to achieve these limits suggests that any systematic low-frequency radio study of the auroral environment of substellar neighbours will require substantial dedication of currently available resources, or an improvement in the low-frequency capability not currently foreseen in upcoming facilities such as the Square Kilometre Array.

\section*{Acknowledgements}
BB acknowledges financial support from the European Commission in the form of a Marie Curie International Outgoing Fellowship (PIOF-GA-2013- 629435). MJH \& WLW acknowledge support from the UK's Science and Technology Facilities
Council [grant number ST/M001008/1]. SLC is supported by the College of Science and Engineering at the University of Leicester. SPL acknowledges support from the UK's Science and Technology Facilities
Council [grant number ST/M001350/1].
\bibliographystyle{mn2e}
\bibliography{refs}

\appendix
\section{Images}
Here we show LOFAR images of the fields around the brown dwarfs. These
$1000 \times 1000$ arcsec images are a small fraction of the total
field of view imaged by LOFAR.

\begin{figure*}
\includegraphics[width=\textwidth]{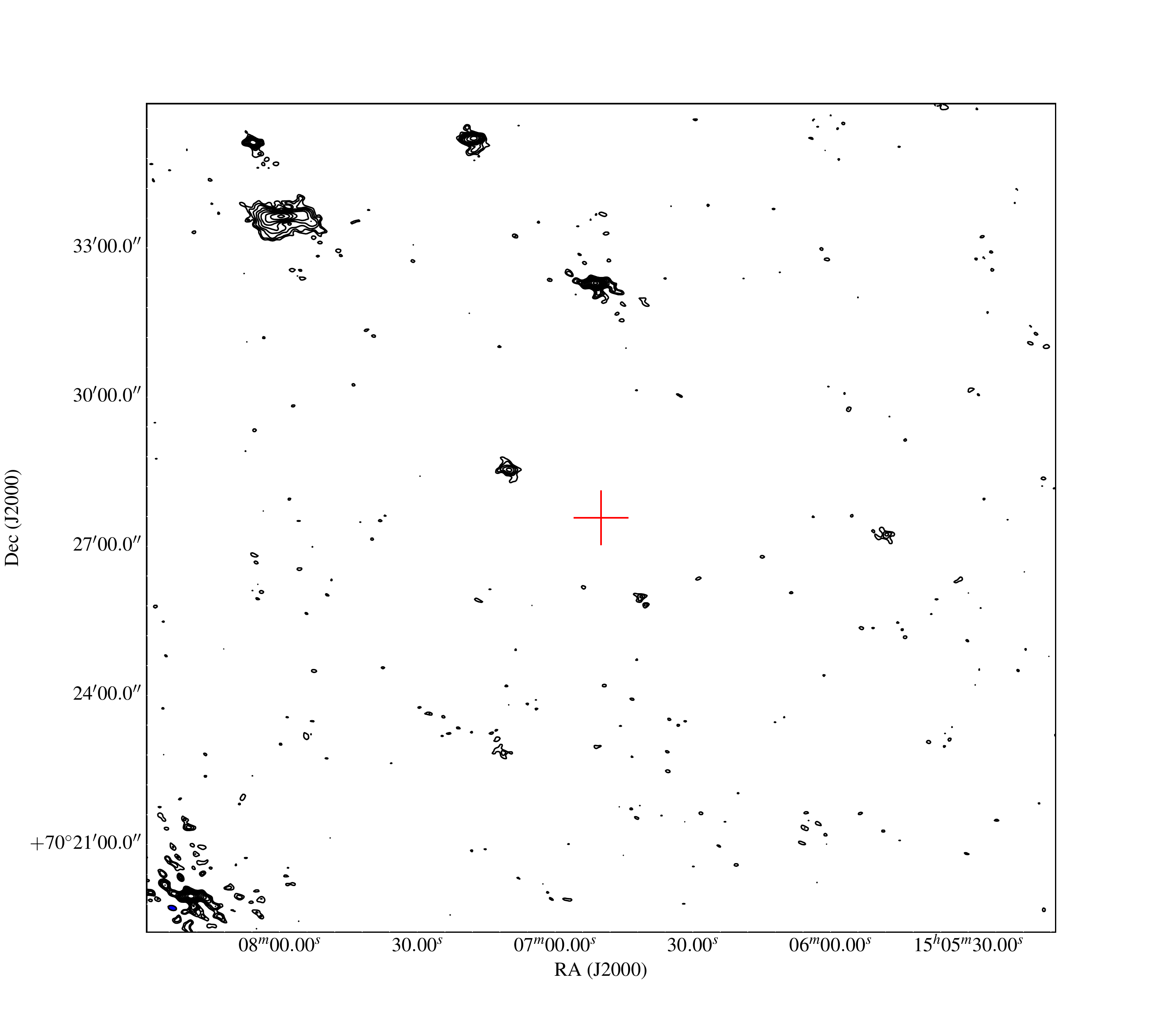}
\caption{The field around WISE1506. A red cross marks the position of
  the target. Rms noise at this position is 240
  $\mu$Jy beam$^{-1}$; contours are at $3 \times 240 \times
  (1,\sqrt{2},2,\dots)$ $\mu$Jy beam. The resolution of the image is
  $9.3 \times 5.5$ arcsec, shown by a blue ellipse in the bottom
  left-hand corner.}
\label{fig:w1506}
\end{figure*}

\begin{figure*}
\includegraphics[width=\textwidth]{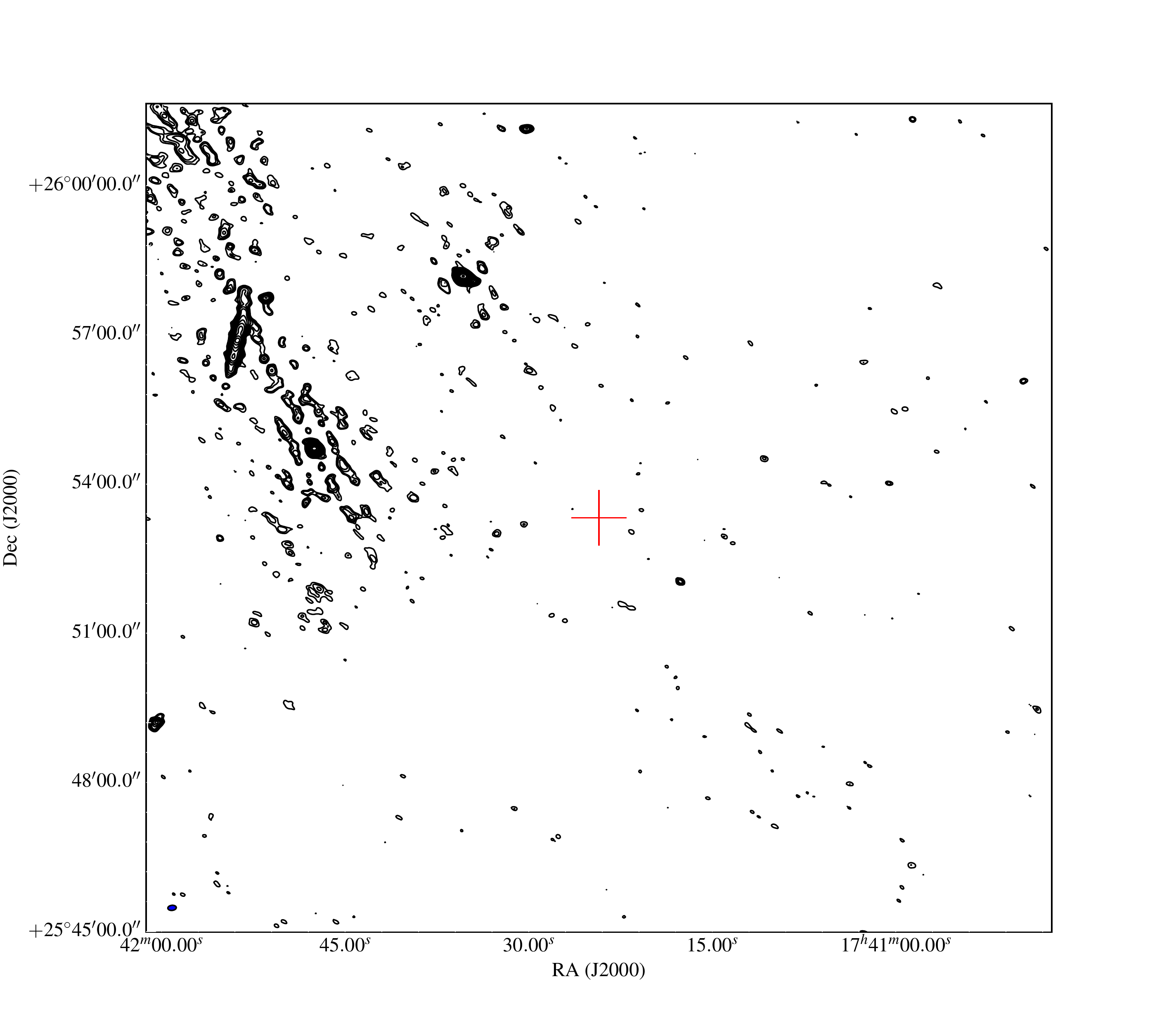}
\caption{The field around WISE1741. As Figure \ref{fig:w1506}, but the
rms noise at this position is 290 $\mu$Jy beam$^{-1}$ and the
resolution $9.4 \times 6.3$ arcsec. The extended source referred to in
the text can be seen to the NE.}
\label{fig:w1741}
\end{figure*}

\begin{figure*}
\includegraphics[width=\textwidth]{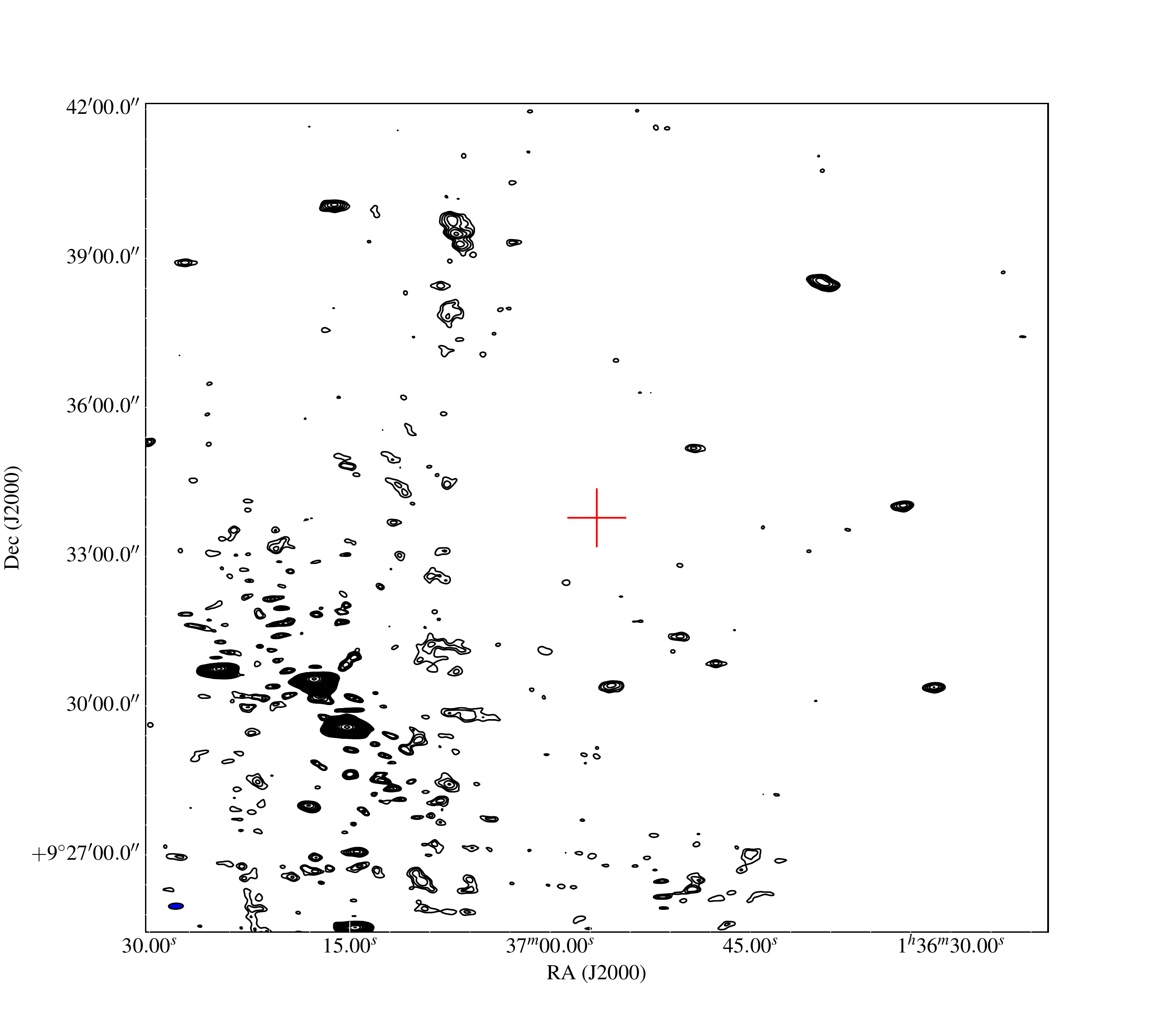}
\caption{The field around SIMP0136. As Figure \ref{fig:w1506}, but the
rms noise at this position is 220 $\mu$Jy beam$^{-1}$ and the
resolution $16.1 \times 6.7$ arcsec. In the SE of the image dynamic
range limitation around the bright source 4C +09.06 (4.7 Jy) can be seen.}
\label{fig:s0136}
\end{figure*}

\end{document}